\documentstyle[12pt,prl,manuscript,psfig,aps,amssymb]{revtex}
 
\begin{document} 
\draft 
\preprint{}

\title{Crystallographic Restrictions for Colour Lattices \\
with Modular Sublattices}

\author{Leonid G. Fel}
\address{School of Physics and Astronomy \\Raymond and Beverly Sackler 
Faculty of Exact Sciences,\\ Tel Aviv University,  
Tel Aviv 69978, Israel\\}

\date{\today}

\maketitle

\def\be{\begin{equation}} 
\def\ee{\end{equation}} 
\def\p{\prime}


\subsection*{Abstract}

The {\em d} -- dimensional $n$ -- colour lattice ${\bf \Bbb{L}}^d$ with
modular sublattices are studied, when the only one crystallographic 
type of sublattices does exist and the only one of the colours occupies
a sublattice, which is still invariant under $k$--fold rotation $C_k$.
Such kind of colouring always preserves an equal fractions of the 
colours composed ${\bf \Bbb{L}}^d$. The $n$ -- colour lattice with
modular sublattices allow to exist the crystallographic rotations $C_k$ 
for $k=p^r,\;r\geq 1$ and $n\leq p$, where $p$ is a prime number.


\noindent 

\widetext 
\narrowtext 
\newpage
\section{Introduction}

One of the fascinating results in the {\em d} -- dimensional (${\em d}$D) crystallography 
is the Hermann theorem \cite{herm49} on the {\it crystallographic 
restriction} ({\sf CR}): in ${\em d}$D -- lattice ${\bf \Bbb{L}}^d$ 
there are allowed the crystallographic rotations $C_k$ of all $k$ which satisfy 
$d\geq \Psi(k)$, where $\Psi(k)$ is the Euler totient function. Due to Schl\"afli 
\cite{schl66}, a lattice ${\bf \Bbb{L}}^{2d+1}$ of odd dimension always possesses 
a rotation axis $C_k$ if a lattice ${\bf \Bbb{L}}^{2d}$ possesses this axis, that 
reflects an even values of the totient function $\Psi(k)$.

Recent progress in the problem of effective conductivity in 2D three -- component 
regular composites \cite{fel20} poses a question about {\sf CR} in a plane 
{\em tiling} perfectly coloured with $n$ different colours of equal fractions.  
This question is concerned with another Hermann theorem \cite{herm34} about material 
tensors of $r$--th rank operating in media with an inner symmetry that includes a 
rotation axis of the highest order $l$. namely, the existence of the rotation axis 
$C_l$ of order $l> r$ makes the $r$--th rank tensor isotropic. 
There are two other sorts of colour lattices which are of high importance in 
physics of condensed matter. These are the lattices with colour {\em edges},  
which appear in the {\it bond percolation problem} 
\cite{essam80} and the lattices with colour {\em nodes} which are related to the
{\it site percolation problem} in the theory of phase transition \cite{wu82}. 

\begin{figure}[h]
\centerline{\psfig{figure=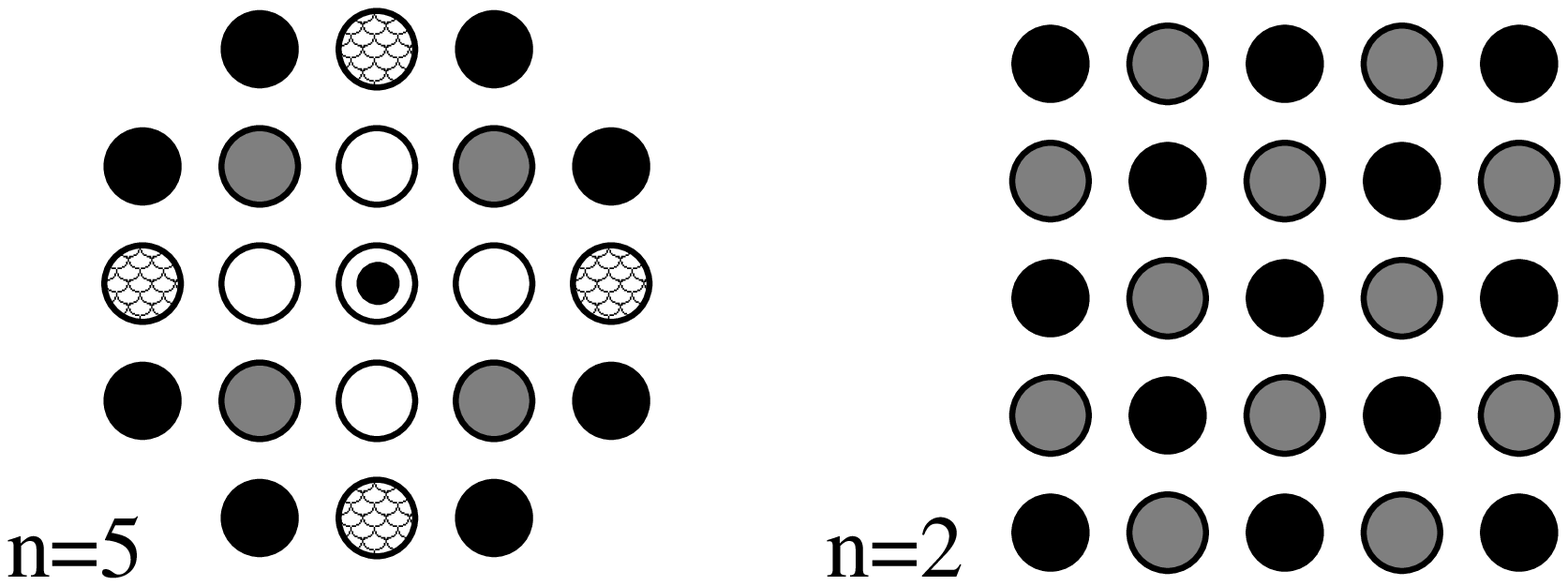,height=5cm,width=14cm}}
\vspace{.5cm}
\caption{}
\label{c5col}
\end{figure}

The last kind of colour lattices is most simplest from crystallographic standpoint.
Recently the group -- theoretical methods were successfully used here to find 
the different possibilities to colour the lattice sites \cite{ples96}, 
\cite{baak97}. We have in mind the coincidence problem for the colouring a 
lattice possessed a $k$ -- fold rotation with finitely many 
pairwise distinct colours in such a way that the colour version is still 
crystallographic: one of the colours occupies a sublattice which is still 
invariant under $k$ -- fold rotation, while the other colours label the cosets. 
In other words a problem was reduced to the question: how many sublattices with 
$k$ -- fold rotation of ${\mathbb L}^d$ of a given index $n$ do exist. In 
2D lattices the classification of such coincidence is the same as the 
classification of the colour lattices with rotational symmetry 
\cite{sch80}.  For such 
kind of colouring the obtained results \cite{ples96}, \cite{baak97} can be 
interpret as a generalization of Hermann theorem for colour lattices, e.g. in 
the 2D $n$ -- collour lattices the following crystallographic 
rotations $C_k$ are allowed 
\begin{eqnarray}
C_4\;\rightarrow\;n=1,2,4,5,8,9,10,13,16,\;...\;,\;\;\;\;
C_6\;\rightarrow\;n=1,3,4,7,9,12,13,16,\;...\;\;.\nonumber
\end{eqnarray}
For indecomposable point groups in higher dimensions no nontrivial 
orthogonal transformation commutes with all symmetries so there is no 
longer this equivalence. 
One can see however that even in the 2D -- case the colouring rules 
defined 
in \cite{ples96}, 
\cite{baak97} admit both the lattices with a permutation invariance of 
different sublattices as well as the lattices where this property is absent 
(see Fig. \ref{c5col}). 
Moreover such colouring admits a non--equal fractions of
the colours composed the entire lattice.

A new algebraic approach \cite{felgaus02} to the effective isotropic 
conductivity problem of 2D infinite non -- random composite has shown the 
existence of self -- dual algebraic functions $\lambda_n(x_i)$ which are
invariant upon the action of full permutation group $S_n$. The regular arrangement 
of $n$ colour nodes at the plane, which are corresponding to distinct 
constituents, is restricted by growing $n$: the existence of rotation symmetry
with $C_k,\;k\geq 3$ preserving the permutation invariance of the equal fraction 
constituents disappear when $n>4$. By other hand the mentioned properties can be 
restored if we admit the quasi--crystalline arrangement of the nodes at the plane
via their projection from the high--dimensional colour lattice, which possessed
a high--fold rotation axis.

This motivates us to study such kind of colouring where the only one crystallographic
type of sublattices does exist: any two sublattices ${\bf \Bbb{L}}^d_{q_1}(n)$ 
and ${\bf \Bbb{L}}^d_{q_2}(n)$, which belong to the distinct colours $q_1$ and $q_2$ 
respectively, are isomorphic ${\bf \Bbb{L}}^d_{q_1}(n)\leftrightarrow
{\bf \Bbb{L}}^d_{q_2}(n)$ via translational motion by a distance and in a direction 
equal to one of the basic vectors. An isomorphism preserves the equal fractions of 
the colours in the unit cell. Such sublattices will be called {\it modular} due to their 
relation to the modular equations.

The colour lattices, which we are going to study in the present paper, give 
rise to another version of Hermann theorem: what rotational axes are survived 
after colouring the nodes of the lattice ${\bf \Bbb{L}}^d$. In the case 
$d=2$ an answer was found in \cite{fel20a} 
where an existence of the crystallographic rotations with $k=2,3,4$ for 2 -- colour 
lattices and $k=3$ for 3 -- colour lattices was shown. This means  that the 
isotropic tensor of second rank, like conductivity ${\widehat \sigma}$, 
exists in the plane media regularly composed of not more than 3 sort of nodes.

The approach which we exploit here is quite simple: starting from the 
admitted axes $C_k$ we will check which of them are survived after colouring the 
lattice in $n$ colours.

\section{Two -- dimensional Lattices}
Before going to {\sf CR} in the colour lattices we recall its derivation for 2D 
unicolour ($n=1$) lattice. A plane lattice ${\bf \Bbb{L}}^2$ is arranged as a set 
of periodic sequences of nodes defined by vectors ${\bf d_2}\in {\bf \Bbb{L}}^2$ 
\begin{eqnarray}
{\bf d_2}=m_1 {\bf e_1}+m_2 {\bf e_2},\;\;
\langle{\bf e_1},{\bf e_2}\rangle=\cos 
\psi,\;\;\psi\neq 0,\pi,\nonumber
\end{eqnarray} 
where $m_1, m_2\in \bf \Bbb{Z}$, and ${\bf e_1} , {\bf e_2}$ are two 
noncollinear unit vectors, called lattice basis. Such set contains infinite 
number of points. Any translational motion of this lattice by a distance and in 
a direction equal to 
one of the basic vectors superimposes the points of the set upon their original
positions, so that the new state of the set is indistinguishable from the old one. 

The rotational part ${\widehat R}$ of every crystallographic symmetry operation
has a representation through a unimodular $2\times2$ matrix with integral entries. 
It is simply to establish this matrix for counter--clockwise isometric 
rotation
${\widehat R}$ of the order $k=2\pi/\phi_k\;,\;\left({\widehat R}_k\right)^k={\widehat I_2}$ :
\begin{eqnarray}
{\widehat R}_k=\frac{1}{\sin \psi_k}
\left(\begin{array}{cc}
\sin (\psi_k-\phi_k)& -\sin \phi_k \\
\sin \phi_k & \sin (\psi_k+\phi_k) 
\end{array}\right)\;,\;\;\det {\widehat R}_k=1\;,
\label{l22}
\end{eqnarray}
where ${\widehat I_2}$ is a $2\times2$ identity matrix. The admitted values of 
$\psi_k$ and $\phi_k$ are exhausted by the following list
\begin{eqnarray}
&&C_1:\phi_1=2\pi,\;\;{\widehat R}_1={\widehat I}\;,\;\;
C_2:\;\phi_2=\pi\;,\;\;{\widehat R}_2=-{\widehat I}\;,\nonumber\\
&&C_k:\phi_k=\psi_k=\frac{2\pi}{k},\;{\widehat R}_k=
\left(\begin{array}{rr}
0  &  -1 \\
1 & a_k \end{array}\right),\;\;
a_k=\left\{\begin{array}{r}
-1\;,\;k=3\\
0\;,\;k=4\\
1\;,\;k=6\end{array}\right.
\label{l23}
\end{eqnarray}
We will define now a colour plane lattice ${\bf \Bbb{L}}^2(n)$ as a union of 
$n$ {\em modular} sublattices ${\bf \Bbb{L}}^2_q(n), q=0,1,...,n-1$, each of 
them has a separate colour. The construction of $q$--th sublattice ${\bf 
\Bbb{L}}^2_q(n)$ is the following
\begin{eqnarray}
&&{\bf \Bbb{L}}^2_q(n)=\bigcup_{m_i^{(q)}\in M_{q,n}} 
{\bf d_2}^{q,n},\;\;\;0\leq q\leq n-1\;,\label{l24}\\
&&{\bf d_2}^{q,n}=m_1^{(q)} {\bf e_1}+m_2^{(q)} {\bf e_2}\;,\;\;
m_1^{(q)}+m_2^{(q)}=q\;\bmod(n)\nonumber
\end{eqnarray}
and all cyclic rotations $\left({\widehat R}_k\right)^t$ preserve the invariance
of sublattice ${\bf\Bbb{L}}^2_q(n)$
\begin{eqnarray}
\left({\widehat R}_k\right)^t{\bf d_2}^{q,n}
\in {\bf\Bbb{L}}^2_q(n)\;,\;\;0\leq t\leq k-1\;.
\label{l24a}
\end{eqnarray}
Here $M_{q,n}$ denotes a set of integer solutions $m_1^{(q)},m_2^{(q)}$
of the system of $k$ linear modular equations (\ref{l24}, \ref{l24a}). 
Such definition of ${\bf\Bbb{L}}^2(n)$ has one important corollary: 
if ${\bf\Bbb{L}}^2(n)$ is invariant upon the isometric rotation ${\widehat R}_k$ 
then there exist $(n-1)$ -- colour lattice ${\bf\Bbb{L}}^2(n-1)$ which is also
invariant under rotation ${\widehat R}_k$. Indeed, let us delete one modular
sublattice ${\bf\Bbb{L}}^2_{1}(n)$ from the entire ${\bf\Bbb{L}}^2(n)$. This does 
not affect the existence of the other $n-1$ relations (\ref{l24}, \ref{l24a}) and
therefore the ${\bf\Bbb{L}}^2(n-1)$ -- lattice remains to be invariant.
In fact, this is true for any number of deleted sublattices. Thus, we conclude
that if the ${\bf\Bbb{L}}^2$ -- lattice could be coloured in $n$ colours, then
it also could be coloured in any number smaller than $n$. We denote such 
maximal number of colours in the plane lattice with $N_k^2$.

Let us consider now a plane lattice ${\bf \Bbb{L}}^2$ spanned on the basic vectors 
${\bf e_1},{\bf e_2}$ 
\begin{eqnarray}
{\bf e_1}=\left(\begin{array}{c}
1\\0\end{array}\right),\;
{\bf e_2}=\left(\begin{array}{c}
0\\1\end{array}\right)\;\nonumber
\end{eqnarray}
and possessed a rotation axis $C_k$. We mark the lattice in $n$ colours in 
accordance with (\ref{l24}), (\ref{l24a}). Acting with an isometric rotation 
${\widehat R}_k$ (\ref{l23}) onto the entire colour lattice ${\bf 
\Bbb{L}}^2(n)$
\begin{eqnarray}
{\widehat R}_k\; {\bf e_1}={\bf e_2}\;,\;\;
{\widehat R}_k\; {\bf e_2}=-\;{\bf e_1}+a_k\;{\bf e_2}\nonumber
\end{eqnarray}
we act by the same on all sublattices ${\bf \Bbb{L}}^2_q(n)$ 
\begin{eqnarray}
{\widehat R}_k\; {\bf d_2}^{q,n}=-m_2^{(q)}\;{\bf e_1}+
\left(m_1^{(q)}+a_k m_2^{(q)}\right){\bf e_2}\;.
\label{l26}
\end{eqnarray}
Let us consider the corresponding system of two modular equations $t=0,1$
\begin{eqnarray}
&&m_1^{(q)}+m_2^{(q)}=q\;\bmod(n)\;,\;\;q=0,...,n-1 \label{l27}\\
&&m_1^{(q)}+(a_k-1) m_2^{(q)}=q\;\bmod(n)\;.\label{l27a}
\end{eqnarray}
Both equations (\ref{l27}), (\ref{l27a}) are coincident when the equality  
$(2-a_k)=0\;\bmod(n)$ is valid. This leads to the conclusion
\begin{equation}
N_k^2=2-a_k\;\;\longrightarrow\;\;N_3^2=3\;,\;N_4^2=2\;,\;N_6^2=1
\label{l28}
\end{equation}
in accordance with \cite{fel20a}. It is simply to convince that the other 
$k-2$ modular equations (\ref{l24a}) do not contradict with (\ref{l28}). 

\begin{figure}[h]
\centerline{\psfig{figure=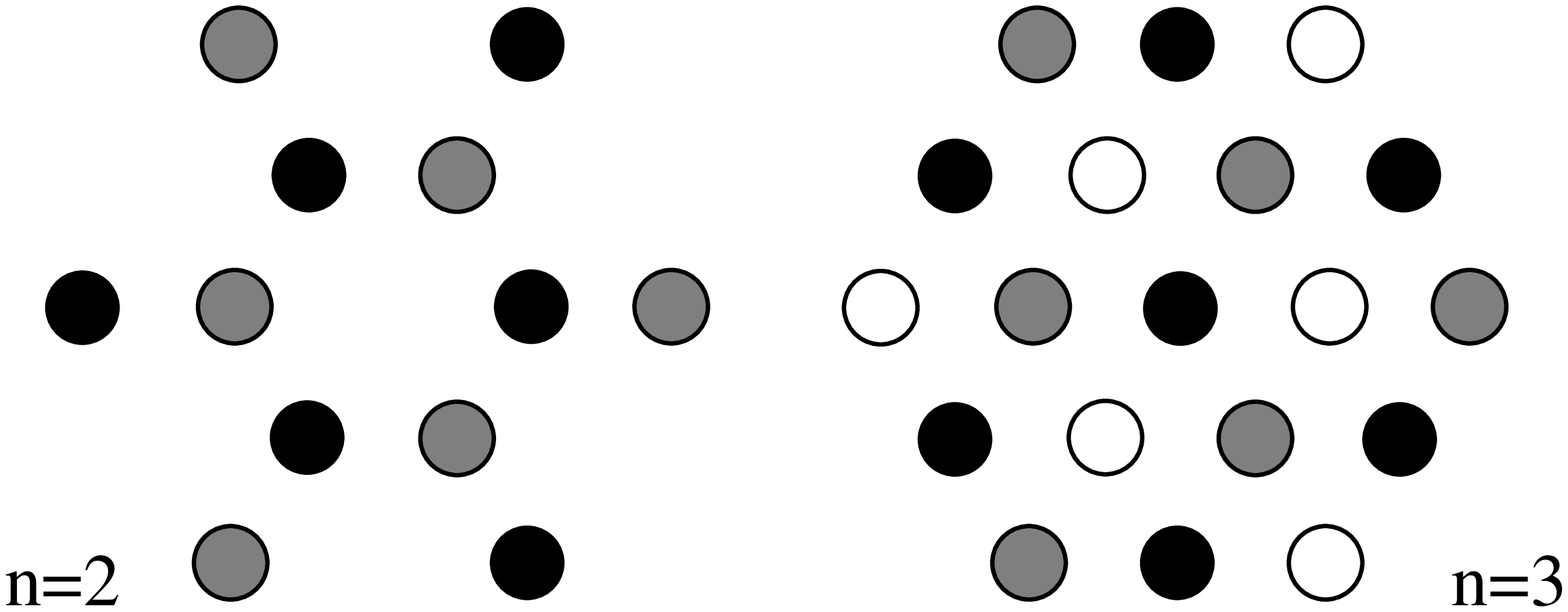,height=5.5cm,width=14cm}}
\vspace{.5cm}
\caption{}
\label{c3col}
\end{figure}

At Figure \ref{c3col} we present two different 
colourings of the plane lattice with modular sublattices which preserve the $C_3$ 
rotation axis. The similar treatment for two -- fold axis $C_2$ is trivial
\begin{eqnarray}
{\widehat R}_2\; {\bf e_i}=-{\bf e_i}\;,\;\;
{\widehat R}_2\; {\bf d_2}^{q,n}=-{\bf d_2}^{q,n}\;,\nonumber
\end{eqnarray}
that yields
\begin{eqnarray}
m_1^{(q)}+m_2^{(q)}=\pm q\;\bmod(n)\;\longrightarrow\;\;
2\left(m_1^{(q)}+m_2^{(q)}\right)=0\;\bmod(n)\;\longrightarrow\;\;
N_2^2=2\;.
\label{twofold}
\end{eqnarray}
In 3D lattice the numbers of colours $N_k^3$ which preserve the modular 
invariance (\ref{l24}) remain the same like in 2D lattice.

\section{High -- dimensional lattices and unimodular matrix representations}
In this Section we consider {\sf CR} in high -- dimensional colour lattices 
${\bf \Bbb{L}}^d(n)$ with modular sublattices. We define such lattice in a way 
similar to (\ref{l24}): it is a union of $n$ {\em modular} sublattices 
${\bf \Bbb{L}}^d_q(n), q=0,1,...,n-1$, each of them has a separate colour. 
The construction of $q$--th sublattice ${\bf \Bbb{L}}^d_q(n)$ is the following
\begin{eqnarray}
&&{\bf \Bbb{L}}^d_q(n)=\bigcup_{m_i^{(q)}\in M_{q,n}}
{\bf d}_d^{q,n},\;\;\;0\leq q\leq n-1\;,\label{n24}\\
&&{\bf d}_d^{q,n}=\sum_{i=1}^{d}m_i^{(q)} {\bf e_i}\;,\;\;\;
\sum_{i=1}^{d}m_i^{(q)}=q\;\bmod(n)\;,\;\;\;0\leq q\leq n-1\;,\nonumber
\end{eqnarray}
where the basic vectors read
\begin{eqnarray}
{\bf e_1}=\left(\begin{array}{c}
1\\0\\...\\...\\0\\0\end{array}\right),\;
{\bf e_2}=\left(\begin{array}{c}
0\\1\\...\\...\\0\\0\end{array}\right),\;\;
.\;.\;.\;\;,\;
{\bf e_d}=\left(\begin{array}{c}
0\\0\\...\\...\\0\\1\end{array}\right).\nonumber
\end{eqnarray}
A straightforward way to built out the linear modular equations using the 
matrix representations for every axial rotation $C_k$ in ${\bf\Bbb{L}}^d$
is cumbersome due to rapid growth of both a number $k$ of equations and a 
dimension $\Psi(k)$ of irreducible representations. To avoid such algebra
we will make use of a special property of axial rotation $C_p$, where $p$ 
is a prime number. This is a unique rotation, all the powers of which 
have irreducible representations. Such sort of rotations arranges a natural
basis to utilize all other rotations $C_k$ with non -- prime $k$.

\subsection{Rotation symmetry $C_k,\;k=p$ .}
\label{subsect1}
This symmetry operation has an irreducible representation ${\widehat R}_p$ 
through a unimodular companion matrix \cite{lang87} $\det{\widehat R}_p=1,\;
\dim {\widehat R}_p=p-1$
\begin{eqnarray}
{\widehat R}_p=
\left(\begin{array}{rrrrrrrr}
0 & 0 & 0 & ... & 0 & 0 & -1\\
1 & 0 & 0 & ... & 0 & 0 & -1\\
0 & 1 & 0 & ... & 0 & 0 & -1\\
... & ... & ... & ... & ...& ...& ...\\
0 & 0 & 0 & ... & 1 & 0 & -1\\
0 & 0 & 0 & ... & 0 & 1& -1\end{array}\right),\;
\left({\widehat R}_p\right)_{i,j}=\left\{\begin{array}{l}
\;\;\;1\;,\;\mbox{if}\;j=i-1\;,\;2\leq i\leq p-1\;,\\
-1\;,\;\mbox{if}\;j=p-1\;,\\
\;\;\;0\;,\;\mbox{otherwise}\;,\end{array}\right.
\label{xaxa1}
\end{eqnarray}
where $\left({\widehat R}_p\right)^p={\widehat I}_{\pi},\;
\dim {\widehat I}_{\pi}=p-1$ and ${\widehat I}_{\pi}$ is an identity matrix. 
In other words ${\widehat R}_p$ is a generator of Abelian group $C_p$ over the integer 
numbers ${\bf \Bbb{Z}}$. All the powers of ${\widehat R}_p$ are also irreducible matrices
\begin{eqnarray}
\left({\widehat R}_p\right)^t_{i,j}=\left\{\begin{array}{l}
\;\;\;1\;,\;\mbox{if}\;j=p+i-t\;,\;1\leq i\leq t-1\;,\\
-1\;,\;\mbox{if}\;j=p-t\;,\\
\;\;\;1\;,\;\mbox{if}\;j=i-t\;,\;t+1\leq i\leq p-1\;,\\
\;\;\;0\;,\;\mbox{otherwise}\;.\end{array}\right.
\label{xaxa2}
\end{eqnarray}
E.g. 
\begin{eqnarray}
{\widehat R}_5=
\left(\begin{array}{rrrr}
0 & 0 & 0 & -1\\
1 & 0 & 0 & -1\\
0 & 1 & 0 & -1\\ 
0 & 0 & 1 &- 1\end{array}\right),\;
\left({\widehat R}_5\right)^2=
\left(\begin{array}{rrrr}
0 & 0 & -1& 1\\
0 & 0 & -1& 0\\
1 & 0 & -1& 0\\
0 & 1 & -1 & 0\end{array}\right),\;.\;.\;.\;,\;
\left({\widehat R}_5\right)^4=
\left(\begin{array}{rrrr}
-1& 1& 0 & 0\\ 
-1& 0& 1 & 0\\
-1& 0& 0 & 1\\
-1& 0& 0 & 0\end{array}\right).
\label{xaxa2b}
\end{eqnarray}
Let us present its action on the basic vectors ${\bf e_i}$
\begin{eqnarray}
{\widehat R}_p\;{\bf e_{p-1}}=-\sum_{j=1}^{p-1}{\bf e_j}\;,\;\;
{\widehat R}_p\;{\bf e_j}={\bf e_{j+1}}\;,\;\;j=1,...,p-2\;.
\label{xaxa3}
\end{eqnarray}
or more generally 
\begin{eqnarray}
&&\left({\widehat R}_p\right)^t\;{\bf e_{p-t-j}}={\bf 
e_{p-j}}\;,\;\;j=1,...,p-t-1\;,\nonumber\\
&&\left({\widehat R}_p\right)^t\;{\bf e_{p-t}}=
-\sum_{i=1}^{p-1}{\bf e_i}\;,\;\;t=1,...,p-1\;,\nonumber\\
&&\left({\widehat R}_p\right)^t\;{\bf e_{p-t+j}}={\bf 
e_j}\;,\;\;j=1,...,t-1\;.
\label{xaxa4}
\end{eqnarray}
Then the action on a generic vector ${\bf d}_{p-1}^{q,n}$ is
\begin{eqnarray} 
\left({\widehat R}_p\right)^t {\bf d}_{p-1}^{q,n}=
\sum_{j=t+1}^{p-1} m_{j-t}^{(q)} {\bf e_j}-
m_{p-t}^{(q)}\sum_{j=1}^{p-1}{\bf e_j}+\sum_{j=1}^{t-1}m_{p-t+j}^{(q)}{\bf 
e_j}
\nonumber
\end{eqnarray}
i.e.
\begin{eqnarray} 
{\widehat R}_p\;{\bf d}_{p-1}^{q,n}=\sum_{j=2}^{p-1} m_{j-1}^{(q)} {\bf 
e_j}-m_{p-1}^{(q)}\sum_j^{p-1}{\bf e_j}\;,\;\;
\left({\widehat R}_p\right)^2 {\bf d}_{p-1}^{q,n}=
\sum_{j=3}^{p-1} m_{j-2}^{(q)}{\bf e_j}-m_{p-2}^{(q)}\sum_{j=1}^{p-1}{\bf 
e_j}+ m_{p-1}^{(q)}{\bf e_1},\mbox{{\it etc}}.
\nonumber
\end{eqnarray}
The modular equations are the following
\begin{eqnarray}
&&\sum_{j=1}^{p-1} m_{j}^{(q)}=q\;\bmod(n)\;,\;\;q=0,...,n-1\;,\label{xaxa5} \\
&&\sum_{j=1}^{p-1} m_{j}^{(q)}-p\;m_i^{(q)}=q\;\bmod(n)\;,\;\;
i=1,...,p-1\;.\nonumber
\end{eqnarray}
They are coincident if $p=0\;\bmod(n)$. Hence it follows $N_p^{p-1}=p$.
\subsection{Rotation symmetry $C_k,\;k=p_1 p_2,\;p_1\neq p_2$ .}
\label{subsect2}
The case $k=p_1 p_2,\;p_1\neq p_2$ presents another situation, when the 
modular equations do not have any solutions. First we show this in the trivial
case $k=2p,\;p\neq 2$. The irreducible representation ${\widehat 
R}_{2p}=-{\widehat R}_p$ for axial rotation $C_{2p}$ leads finally to the 
modular equations 
\begin{eqnarray}
&&\sum_{j=1}^{p-1} m_{j}^{(q)}=\pm q\;\bmod(n)\;,\;\;q=0,...,n-1\;,
\label{xaxa6} \\
&&\sum_{j=1}^{p-1} m_{j}^{(q)}-p\;m_i^{(q)}=\pm q\;\bmod(n)\;,\;\;
i=1,...,p-1\;.\nonumber
\end{eqnarray}
or, more briefly, 
$$
2\;\sum_{j=1}^{p-1} m_{j}^{(q)}=p\;m_i^{(q)}=0\;\bmod(n)\;,
$$
which do not have solutions for any modulus $n$: $N_{2 p}^{p-1}=1$ . 

The generic case $p_1\neq p_2$ leads to two equivalent irreducible 
representations through a $(p_1-1)(p_2-1) - \dim$ unimodular matrices  
${\widehat A}_{p_1 p_2}={\widehat R}_{p_1}\otimes {\widehat R}_{p_2}$ and 
${\widehat B}_{p_1 p_2}={\widehat R}_{p_2}\otimes {\widehat R}_{p_1}$, where $\otimes$ 
denotes a tensorial product. It is convenient to present they as block --  matrices with 
companion matrix entries
\begin{eqnarray}
{\widehat A}_{p_1 p_2}=
\left(\begin{array}{cccccccc}
{\widehat 0}_{\pi_1} & {\widehat 0}_{\pi_1} & {\widehat 0}_{\pi_1} & ... & {\widehat 0}_{\pi_1} & 
{\widehat 0}_{\pi_1} & -{\widehat R}_{p_1}\\
{\widehat R}_{p_1} & {\widehat 0}_{\pi_1} & {\widehat 0}_{\pi_1} & ... & {\widehat 0}_{\pi_1} & 
{\widehat 0}_{\pi_1} & -{\widehat R}_{p_1}\\
{\widehat 0}_{\pi_1} & {\widehat R}_{p_1} & {\widehat 0}_{\pi_1} & ... & {\widehat 0}_{\pi_1} & 
{\widehat 0}_{\pi_1} & -{\widehat R}_{p_1}\\
... & ... & ... & ... & ... & ...& ...\\
{\widehat 0}_{\pi_1} & {\widehat 0}_{\pi_1} & {\widehat 0}_{\pi_1} & ... & {\widehat R}_{p_1} & 
{\widehat 0}_{\pi_1} & -{\widehat R}_{p_1}\\
{\widehat 0}_{\pi_1} & {\widehat 0}_{\pi_1} & {\widehat 0}_{\pi_1} & ... & {\widehat 0}_{\pi_1} & 
{\widehat R}_{p_1} & -{\widehat R}_{p_1}
\end{array}\right),\;
{\widehat B}_{p_1 p_2}=
\left(\begin{array}{cccccccc}
{\widehat 0}_{\pi_2} & {\widehat 0}_{\pi_2} & {\widehat 0}_{\pi_2} & ... & {\widehat 0}_{\pi_2} &
{\widehat 0}_{\pi_2} & -{\widehat R}_{p_2}\\
{\widehat R}_{p_2} & {\widehat 0}_{\pi_2} & {\widehat 0}_{\pi_2} & ... & {\widehat 0}_{\pi_2} &
{\widehat 0}_{\pi_2} & -{\widehat R}_{p_2}\\
{\widehat 0}_{\pi_2} & {\widehat R}_{p_2} & {\widehat 0}_{\pi_2} & ... & {\widehat 0}_{\pi_2} &
{\widehat 0}_{\pi_2} & -{\widehat R}_{p_2}\\
... & ... & ... & ... & ... & ...& ...\\
{\widehat 0}_{\pi_2} & {\widehat 0}_{\pi_2} & {\widehat 0}_{\pi_2} & ... & {\widehat R}_{p_2} &
{\widehat 0}_{\pi_2} & -{\widehat R}_{p_2}\\
{\widehat 0}_{\pi_2} & {\widehat 0}_{\pi_2} & {\widehat 0}_{\pi_2} & ... & {\widehat 0}_{\pi_2} &
{\widehat R}_{p_2} & -{\widehat R}_{p_2}
\end{array}\right),
\nonumber
\end{eqnarray}  
where ${\widehat R}_{p_1}, {\widehat R}_{p_2}$ can be extracted from (\ref{xaxa1}). The zero 
-- matrices ${\widehat 0}_{\pi_1}, {\widehat 0}_{\pi_2}$ have $\dim {\widehat 0}_{\pi_1}=p_1-1,\;\dim {\widehat 0}_{\pi_2}=p_2-1$. 
The successive powers of ${\widehat A}_{p_1 p_2}$ and ${\widehat B}_{p_1 p_2}$ are the 
diagonal block -- matrices
\begin{eqnarray}
\left({\widehat A}_{p_1 p_2}\right)^{p_2 t_2}&=&\mbox{diag}\overbrace{
\left[\left({\widehat R}_{p_1}\right)^{p_2 t_2},
\left({\widehat R}_{p_1}\right)^{p_2 t_2},\;.\;.\;.\;,
\left({\widehat R}_{p_1}\right)^{p_2 t_2}\right]}^{p_2-1}\;,\;\;t_2=1,...,p_1\;,
\label{diag1}\\
\left({\widehat B}_{p_1 p_2}\right)^{p_1 t_1}&=&\mbox{diag}\overbrace{
\left[\left({\widehat R}_{p_2}\right)^{p_1 t_1},
\left({\widehat R}_{p_2}\right)^{p_1 t_1},\;.\;.\;.\;,
\left({\widehat R}_{p_2}\right)^{p_1 t_1}\right]}^{p_1-1}\;,\;\;t_1=1,...,p_2\;.
\nonumber
\end{eqnarray}
We have one more simplification in the last formul{\ae} which comes due to
${\sf gcd}(p_1,p_2)=1$ and 
$\left({\widehat R}_{p_1}\right)^{p_1}={\widehat I}_{\pi_1}$, 
$\left({\widehat R}_{p_2}\right)^{p_2}={\widehat I}_{\pi_2}$. Indeed, when a variable 
$t_2$ runs through the positive integers $1,...,p_1$ then the matrix representation  
$\left({\widehat R}_{p_1}\right)^{p_2 t_2}$ runs actually through the matrices
$\left({\widehat R}_{p_1}\right)^{t_2}$ and two sets of matrices coincide, e.g. for $p_1=5,\;
p_2=7$ we have
$$
\left({\widehat R}_5\right)^{7}=\left({\widehat R}_5\right)^2,\;
\left({\widehat R}_5\right)^{14}=\left({\widehat R}_5\right)^4,\;
\left({\widehat R}_5\right)^{21}=\left({\widehat R}_5\right)^1,\;
\left({\widehat R}_5\right)^{28}=\left({\widehat R}_5\right)^3,\;
\left({\widehat R}_5\right)^{35}=\left({\widehat R}_5\right)^5.
$$
The same is valid for the running variable $t_1$ and matrices 
$\left({\widehat R}_{p_2}\right)^{p_1 t_1}$, i.e. 
\begin{eqnarray}
\left\{\left({\widehat R}_{p_1}\right)^{p_2 t_2}\right\}\equiv
\left\{\left({\widehat R}_{p_1}\right)^{t_2}\right\}\;,\;\;
\left\{\left({\widehat R}_{p_2}\right)^{p_1 t_1}\right\}\equiv
\left\{\left({\widehat R}_{p_2}\right)^{t_1}\right\}\;. \label{sets1}
\end{eqnarray}
Thus the diagonal matrices (\ref{diag1}) can be rewritten as following
\begin{eqnarray}
\left({\widehat A}_{p_1 p_2}\right)^{p_2 t_2}&=&\mbox{diag}\overbrace{
\left[\left({\widehat R}_{p_1}\right)^{t_2},
\left({\widehat R}_{p_1}\right)^{t_2},\;.\;.\;.\;,
\left({\widehat R}_{p_1}\right)^{t_2}\right]}^{p_2-1}\;,\;\;t_2=1,...,p_1\;,
\label{diag2}\\
\left({\widehat B}_{p_1 p_2}\right)^{p_1 t_1}&=&\mbox{diag}\overbrace{
\left[\left({\widehat R}_{p_2}\right)^{t_1},
\left({\widehat R}_{p_2}\right)^{t_1},\;.\;.\;.\;,
\left({\widehat R}_{p_2}\right)^{t_1}\right]}^{p_1-1}\;,\;\;t_1=1,...,p_2\;.
\nonumber    
\end{eqnarray}
\noindent
Let us consider the action $\left({\widehat A}_{p_1 p_2}\right)^{p_2 t_2}$ on the 
basic vectors ${\bf e_j}$. It can be decomposed in $p_2-1$ parts 
$s=0,\;.\;.\;.\;,p_2-2$ 
\begin{eqnarray}
&&\left({\widehat R}_{p_1}\right)^{t_2}\;{\bf e_{s(p_1-1)+p_1-t_2-j}}=
{\bf e_{s (p_1-1)+p_1-j}}\;,\;\;t_2=1,...,p_1-1\;,\nonumber\\
&&\left({\widehat R}_{p_1}\right)^{t_2}\;{\bf e_{s(p_1-1)+p_1-t_2}}=
-\sum_{i=1+s(p_1-1)}^{(s+1)(p_1-1)}{\bf e_i}\;,\nonumber\\
&&\left({\widehat R}_{p_1}\right)^{t_2}\;{\bf e_{s(p_1-1)+p_1-t_2+j}}=
{\bf e_{s(p_1-1)+j}}\;,\;\;j=1,...,p_1-t_2-1\;,
\label{xaxa4a}
\end{eqnarray}
The corresponding modular equations are
\begin{eqnarray}
&&\sum_{j=1}^{(p_2-1)(p_1-1)} 
m_{j}^{(q)}=q\;\bmod(n)\;,\;\;q=0,...,n-1\;,\label{xaxa5} \\
&&\sum_{j=1}^{(p_2-1)(p_1-1)} 
m_{j}^{(q)}-p_1\;\sum_{s=0}^{p_2-2}
m_{i+s(p_1-1)}^{(q)}=q\;\bmod(n)\;,\;\;
i=1,...,p_1-1\;.\nonumber
\end{eqnarray}
A similar consideration concerned with the action $\left({\widehat B}_{p_1 p_2}\right)^{p_2 t_2}$ 
leads to the modular equations 
\begin{eqnarray}
&&\sum_{j=1}^{(p_2-1)(p_1-1)}
m_{j}^{(q)}=q\;\bmod(n)\;,\;\;q=0,...,n-1\;,\label{xaxa5a} \\
&&\sum_{j=1}^{(p_2-1)(p_1-1)}
m_{j}^{(q)}-p_2\;\sum_{s=0}^{p_1-2}
m_{i+s(p_2-1)}^{(q)}=q\;\bmod(n)\;,\;\;
i=1,...,p_2-1\;.\nonumber
\end{eqnarray}
Both equations (\ref{xaxa5}) and (\ref{xaxa5a}) give a system
\begin{eqnarray}
p_1\;\sum_{s=0}^{p_2-2}m_{i+s(p_1-1)}^{(q)}=
p_2\;\sum_{s=0}^{p_1-2}m_{j+s(p_2-1)}^{(q)}=0\;\bmod(n)\;,\;\;
\left\{\begin{array}{r}
i=1,...,p_1-1\;,\\
j=1,...,p_2-1\;.\end{array}\right.
\label{xaxa5b}
\end{eqnarray}
which do not have solutions for any modulus $n$:
$N_{p_1 p_2}^{(p_1-1)(p_2-1)}=1$ . In Appendix \ref{appendix} we
present two irreducible representations ${\widehat A}_{15}$, ${\widehat 
B}_{15}$ for $C_{15}$ rotation and give a straightforward solution
for corresponding modular equations.

\subsection{Rotation symmetry $C_k,\;k=p^r,\;r\geq 2$.}
\label{subsect3}
This symmetry operation has an irreducible representation ${\widehat R}_{p^r}$ 
through a unimodular companion matrix \cite{lang87} $\det{\widehat 
R}_{p^r}=1,\;\dim {\widehat R}_{p^r}=p^{r-1} (p-1)$
\begin{eqnarray}
{\widehat R}_{p^r}=
\left(\begin{array}{rrrrrrr}
0 & 0 & 0 & ... & 0 & 0 &-1\\
1 & 0 & 0 & ... & 0 & 0 & 0\\
0 & 1 & 0 & ... & 0 & 0 & 0\\
... & ... &... &... &... &... &... \\
0 & 0 & 0 & ... & 0 & 0 &-1\\
... & ... &... &... &... &... &... \\
0 & 0 & 0 & ... & 0 & 0 &-1\\
... & ... &... &... &... &... &... \\
0 & 0 & 0 & ... & 1 & 0 & 0\\
0 & 0 & 0 & ... & 0 & 1 & 0\end{array}\right),\;
\left({\widehat R}_{p^r}\right)_{i,j}=\left\{\begin{array}{l}
\;\;\;1\;,\;\mbox{if}\;j=i-1\;,\;2\leq i\leq p^{r-1} (p-1),\\
-1\;,\;\mbox{if}\; j=p^{r-1} (p-1)\;,\;i=1+s\;p^{r-1},\\
\;\;\;0\;,\;\mbox{otherwise}\;;\;\;s=0,...,p-2\;.\end{array}\right.
\nonumber
\end{eqnarray}
The ${\widehat R}_{p^r}$ representation has a remarkable property
\begin{eqnarray}
\left({\widehat R}_{p^a}\right)^{p^b}=
{\widehat {\cal R}}_{p^{a-b}}\left({\widehat I}_{p^b}\right)\;,\;\;
\mbox{e.g.}\;\;\;\left({\widehat R}_{p^2}\right)^p=
{\widehat {\cal R}}_p\left({\widehat I}_p\right)\;,
\label{xaxarP1}
\end{eqnarray}
where ${\widehat {\cal R}}_{p^{a-b}}\left({\widehat I}_{p^b}\right)$ is formally
coincided with ${\widehat R}_{p^{a-b}}$ and has the entries of the identity matrix
${\widehat I}_{p^b}$ and zero matrix ${\widehat 0}_{p^b}$ instead of 1 and 0. 
An action of ${\widehat R}_{p^r}$ on the basic vectors read
\begin{eqnarray}
{\widehat R}_{p^r}\;{\bf e_{p^{r-1}(p-1)}}=-\sum_{s=0}^{p-2}{\bf
e_{1+s\;p^{r-1}}}\;,\;\;{\widehat R}_{p^r}\;{\bf e_j}={\bf 
e_{j+1}}\;,\;\;j=1,...,p^{r-1}(p-1)-1\;,
\label{xaxarP0}
\end{eqnarray}
\noindent
The corresponding modular equations could be reduced
\begin{eqnarray}
&&\sum_{j=1}^{p^{r-1}(p-1)} m_j^{(q)}=q\;\bmod(n)\;,\;\;q=0,...,n-1\;;
\;\;\;p\; m_1^{(q)}=p\;m_{p^{r-1}(p-1)}^{(q)}=0\;\bmod(n)\;,\nonumber\\
&&p\sum_{j=1}^2 m_j^{(q)}=p\sum_{j=1}^3 m_j^{(q)}=\; .\;.\;.\; =
p\sum_{j=1}^{p^{r-1}-1}m_j^{(q)}=0\;\bmod(n)\;,\nonumber\\
&&p\sum_{j=p^{r-1}(p-1)-1}^{p^{r-1}(p-1)}m_j^{(q)}=
p\sum_{j=p^{r-1}(p-1)-2}^{p^{r-1}(p-1)}m_j^{(q)}=\; .\;.\;.\; =
p\sum_{j=p^{r-1}(p-2)+2}^{p^{r-1}(p-1)}m_j^{(q)}=0\;\bmod(n)\;,\nonumber\\
&&p\sum_{j=1}^{p^{r-1}} m_{j}^{(q)}=p\sum_{j=1}^{p^{r-1}} m_{j+1}^{(q)}=\;.\;.\;.\;=
p\sum_{j=1}^{p^{r-1}} m_{j+p^{r-1}(p-2)}^{(q)}=0\;\bmod(n)\;.\nonumber
\nonumber
\end{eqnarray}
All equations are coincident if $p=0\;\bmod(n)$. Hence it follows 
 $N_{p^r}^{p^{r-1}(p-1)}=p$.

In Appendix \ref{appendix} we give a straightforward solution
for corresponding modular equations for the rotation axis $C_9$.
\subsection{Rotation symmetry $C_k,\;k=p_1^{r_1}p_2^{r_2},\;p_1\neq p_2,\;
r_1,r_2\geq 1$.}
\label{subsect4}
This symmetry operation has an irreducible representation 
${\widehat R}_{p_1^{r_1}p_2^{r_2}}$ through a unimodular matrix 
 $\det{\widehat R}_{p_1^{r_1}p_2^{r_2}}=1,\;\dim 
{\widehat R}_{p_1^{r_1}p_2^{r_2}}=p_1^{r_1-1}p_2^{r_2-1} (p_1-1)(p_2-1)$.
Its $p_1^{r_1-1}p_2^{r_2-1}$ -- power is a diagonal block -- matrix
\begin{eqnarray}
\left({\widehat R}_{p_1^{r_1}p_2^{r_2}}\right)^{p_1^{r_1-1}p_2^{r_2-1}}=
\mbox{diag}\overbrace{
\left[{\widehat R}_{p_1 p_2},\;{\widehat R}_{p_1 p_2},
\;.\;.\;.\;,\;{\widehat R}_{p_1 p_2}\right]}^{p_1^{r_1-1}p_2^{r_2-1}}
\;,\nonumber
\end{eqnarray}
which decomposes the action of $\left({\widehat R}_
{p_1^{r_1}p_2^{r_2}}\right)^{t\;p_1^{r_1-1}p_2^{r_2-1}},\;t=1,2,...,p_1 
p_2$, on the basic vectors ${\bf e_j},\;j=1,2,...,p_1^{r_1-1}p_2^{r_2-1} 
(p_1-1)(p_2-1)$ into identical actions of 
$\left({\widehat R}_{p_1p_2}\right)^t$ on $p_1^{r_1-1}p_2^{r_2-1}$ sets 
of the basic vectors ${\bf e_j},\;j=1,2,...,(p_1-1)(p_2-1)$. According to Section 
\ref{subsect2} such action does not give rise to the nontrivial solutions of 
corresponding modular equations $N_{k}^{\Psi(k)}=1$, where $k=p_1^{r_1}p_2^{r_2}$.
\subsection{Rotation symmetry $C_k,\;k=\prod_{i=1}^s p_i^{r_i},\;r_i\geq 1$.}
\label{subsect5}
The generic case can be easily reduced to the previous Section \ref{subsect4} and 
consequently does not give rise to the nontrivial solutions of
corresponding modular equations.
\section{Conclusion}
\label{sect5}
In the present paper we considered the colour lattice ${\bf \Bbb{L}}^d(n)$ 
with modular sublattices, when the only one crystallographic type of sublattices 
does exist and the only one of the colours occupies a sublattice, which is still 
invariant under $k$--fold rotation $C_k$. Such kind of colouring always preserves 
an equal fractions of the colours composed ${\bf \Bbb{L}}^d$. The relation 
between the $k$ -- fold of the admitted rotation axis $C_k$ and the number $n$
of colours, which marked every modular sublattice, is given by the following 
formul{\ae}
\begin{eqnarray}
n\leq N_{k}^{\Psi(k)}=\left\{\begin{array}{l}
p\;,\;k=p^r\;,\;r\geq 1\;,\\
1\;,\;k=\prod_{i=1}^{s\geq 2} p_i^{r_i}\;,\;r_i\geq 1\;,\end{array}\right.
\label{herman2}
\end{eqnarray}
which means that the $n$ -- colour lattice with modular sublattices allow 
to exist the crystallographic rotations $C_k$ for $k=p^r,\;r\geq 1$ and 
$n\leq p$, where $p$ is a prime number. A simple corollary of (\ref{herman2})
gives an answer to the question: what is a minimal dimension $d$ of 
the colour lattice ${\bf \Bbb{L}}^d(n)$ with $n$ modular sublattices, which 
possess the rotation axis $C_k\;,\;k\geq n$  
\begin{eqnarray}
d=k-1\;,\;\;\;k=\left\{\begin{array}{l}
n\;,\;\;\mbox{if}\;\;n=p\;,\\
p_{min}=\min\{p_j\;;\;p_j > n\}\;,\end{array}\right.
\label{herman3}
\end{eqnarray}
where $p_{min}$ is a minimal prime number which exceeds $n$. Both formul{\ae} 
(\ref{herman2}), (\ref{herman3}) extend the Hermann theorem of the 
crystallographic restrictions on the colour lattices with modular sublattices.

\section{Acknowledgement}
I'm thankful to A. Johasz and I. Kaganov for useful discussions.

This research was supported in part by grants from the U.S. -- Israel
Binational Science Foundation, the Israel Science Foundation, the Tel Aviv 
University Research Authority, and Gileadi Fellowship program of the Ministry 
of Absorption of the State of Israel.

This paper is dedicated to the memory of Prof. V. L. Indenbom which had
made a significant contribution to the theory of crystalline symmetry.

\newpage
\appendix
\renewcommand{\theequation}{\thesection\arabic{equation}}
\section{$C_{15}$ and $C_9$ rotation axes.}
\label{appendix}
\setcounter{equation}{0} 
First we illustrate the diagonalization (\ref{diag2}) of irreducible representations 
${\widehat A}_{15}$, ${\widehat B}_{15}$ for rotation axis $C_{15}$ by taking their 
successive powers.
\begin{eqnarray}
{\widehat A}_{15}=
\left(\begin{array}{rr}
{\widehat 0}_4 & -{\widehat R}_5\\
{\widehat R}_5 & -{\widehat R}_5\end{array}\right)\;,\;\;
\left({\widehat A}_{15}\right)^{3t}=\mbox{diag}\left[
\left({\widehat R}_5\right)^t,\left({\widehat R}_5\right)^t\right]\;,\;t=1,2,3,4,5
\nonumber
\end{eqnarray}
and 
\begin{eqnarray}
{\widehat B}_{15}=
\left(\begin{array}{rrrr}
{\widehat 0}_2 & {\widehat 0}_2 & {\widehat 0}_2 & -{\widehat R}_3\\
{\widehat R}_3 & {\widehat 0}_2 & {\widehat 0}_2 & -{\widehat R}_3\\
{\widehat 0}_2 & {\widehat R}_3 & {\widehat 0}_2 & -{\widehat R}_3\\
{\widehat 0}_2 & {\widehat 0}_2 & {\widehat R}_3 & -{\widehat
R}_3\end{array}\right)\;,\;\;
\left({\widehat B}_{15}\right)^{5t}=\mbox{diag}\left[
\left({\widehat R}_3\right)^t,\left({\widehat R}_3\right)^t,\left({\widehat
R}_3\right)^t,\left({\widehat R}_3\right)^t\right]\;,\;t=1,2,3.
\nonumber
\end{eqnarray}
Let us give a straightforward solution of modular equations for ${\widehat A}_{15}$ 
irreducible representation
\begin{eqnarray}
&&{\widehat A}_{15}=
\left(\begin{array}{rrrrrrrr}
0 & 0 & 0 & 0 & 0 & 0 & 0 & 1\\
0 & 0 & 0 & 0 &-1 & 0 & 0 & 1\\
0 & 0 & 0 & 0 & 0 &-1 & 0 & 1\\
0 & 0 & 0 & 0 & 0 & 0 &-1 & 1\\
0 & 0 & 0 &-1 & 0 & 0 & 0 & 1\\
1 & 0 & 0 &-1 & -1& 0 & 0 & 1\\
0 & 1 & 0 &-1 & 0 &-1 & 0 & 1 \\
0 & 0 & 1 &-1 & 0 & 0 &-1 & 1\end{array}\right)\;,\;\;\;
\left.\begin{array}{ll}
{\widehat A}_{15}\cdot {\bf e}_1={\bf e}_6\;,
& {\widehat A}_{15}\cdot {\bf e}_5=-{\bf e}_2-{\bf e}_6 \\
{\widehat A}_{15}\cdot {\bf e}_2= {\bf e}_7\;,
& {\widehat A}_{15}\cdot {\bf e}_6=-{\bf e}_3-{\bf e}_7 \\
{\widehat A}_{15}\cdot {\bf e}_3={\bf e}_8\;,
& {\widehat A}_{15}\cdot {\bf e}_7=-{\bf e}_4-{\bf e}_8\\
{\widehat A}_{15}\cdot {\bf e}_4=-\sum_{j=5}^8 {\bf e}_j\;,
& {\widehat A}_{15}\cdot {\bf e}_8=\sum_{j=1}^8 {\bf e}_j\;,
\end{array}\right.\nonumber
\end{eqnarray}
Isometric transformation looks like 
\begin{eqnarray}
{\widehat A}_{15}\cdot {\bf d}_8^{q,n}&=&m_8^{(q)}{\bf e}_1+
\left(m_8^{(q)}-m_5^{(q)}\right){\bf e}_2+
\left(m_8^{(q)}-m_6^{(q)}\right){\bf e}_3+
\left(m_8^{(q)}-m_7^{(q)}\right){\bf e}_4+\nonumber\\
&&\left(m_8^{(q)}-m_4^{(q)}\right){\bf e}_5+ 
\left(m_8^{(q)}+m_1^{(q)}-m_4^{(q)}-m_5^{(q)}\right){\bf e}_6+\nonumber\\
&&\left(m_8^{(q)}+m_2^{(q)}-m_4^{(q)}-m_6^{(q)}\right){\bf e}_7+
\left(m_8^{(q)}+m_3^{(q)}-m_4^{(q)}-m_7^{(q)}\right){\bf e}_8
\nonumber
\end{eqnarray}
Modular equations
\begin{eqnarray}
&&m_1^{(q)}+m_2^{(q)}+m_3^{(q)}+m_4^{(q)}+
m_5^{(q)}+m_6^{(q)}+m_7^{(q)}+m_8^{(q)}=q\;\bmod(n)\;,\;\;q=0,...,n-1
\nonumber\\
&&3\left(m_1^{(q)}+m_2^{(q)}+m_3^{(q)}+m_4^{(q)}\right)=
3\left(m_5^{(q)}+m_6^{(q)}+m_7^{(q)}+m_8^{(q)}\right)=0\;\bmod(n)
\nonumber\\
&&5\left(m_1^{(q)}+m_5^{(q)}\right)=5\left(m_2^{(q)}+m_6^{(q)}\right)=
5\left(m_3^{(q)}+m_7^{(q)}\right)=5\left(m_4^{(q)}+m_8^{(q)}\right)=0\;
\bmod(n)\nonumber\\
&&5\left(m_1^{(q)}-2m_5^{(q)}\right)=5\left(m_5^{(q)}-2m_1^{(q)}\right)=
5\left(m_2^{(q)}-2m_6^{(q)}\right)=5\left(m_6^{(q)}-2m_2^{(q)}\right)=0\;
\bmod(n)\nonumber\\
&&5\left(m_3^{(q)}-2m_7^{(q)}\right)=5\left(m_7^{(q)}-2m_3^{(q)}\right)=
5\left(m_4^{(q)}-2m_8^{(q)}\right)=5\left(m_8^{(q)}-2m_4^{(q)}\right)=0\;
\bmod(n)\nonumber
\end{eqnarray}
are unsolvable for any modulus $n$. 

The corresponding representation ${\widehat R}_9$ looks like 
\begin{eqnarray}
{\widehat R}_9=
\left(\begin{array}{rrrrrr}
0 & 0 & 0 & 0 & 0 & -1\\
1 & 0 & 0 & 0 & 0 &  0\\
0 & 1 & 0 & 0 & 0 &  0\\
0 & 0 & 1 & 0 & 0 & -1\\
0 & 0 & 0 & 1 & 0 &  0\\
0 & 0 & 0 & 0 & 1 &  0
\end{array}\right),\;
\left({\widehat R}_9\right)^3=
\left(\begin{array}{rr}
{\widehat 0}_3 & -{\widehat I}_3\\
{\widehat I}_3 & -{\widehat I}_3
\end{array}\right),\;\; 
\left.\begin{array}{ll}
{\widehat R}_9\cdot {\bf e}_1={\bf e}_2\;,
& {\widehat R}_9\cdot {\bf e}_4={\bf e}_5\;,\\
{\widehat R}_9\cdot {\bf e}_2= {\bf e}_3\;,
& {\widehat R}_9\cdot {\bf e}_5={\bf e}_6\;, \\
{\widehat R}_9\cdot {\bf e}_3={\bf e}_4\;,
& {\widehat R}_9\cdot {\bf e}_6=-{\bf e}_1-{\bf e}_4\;.\end{array}\right.
\nonumber
\end{eqnarray}
It leads to the following modular equations
\begin{eqnarray}
&&\sum_{j=1}^6 m_j^{(q)}=q\;\bmod(n)\;,\;\;q=0,...,n-1\;,\nonumber\\
&&3m_6^{(q)}=3m_1^{(q)}=3\left(m_5^{(q)}+m_6^{(q)}\right)=
3\left(m_1^{(q)}+m_2^{(q)}\right)=
0\;\bmod(n)\;,\nonumber\\
&&3\left(m_4^{(q)}+m_5^{(q)}+m_6^{(q)}\right)=
3\left(m_3^{(q)}+m_4^{(q)}+m_5^{(q)}\right)=0\;\bmod(n)\;,\nonumber\\
&&3\left(m_2^{(q)}+m_3^{(q)}+m_4^{(q)}\right)=
3\left(m_1^{(q)}+m_2^{(q)}+m_3^{(q)}\right)=0\;\bmod(n)\;,\nonumber
\end{eqnarray}
which have a solution $N_9^6=3$.
\newpage

\vspace{1cm}

\newpage
${\bf \;\;FIGURE\;\;CAPTIONS.}$

\vskip 0.5cm

\begin{tabbing}
\= FIG.1. \hspace{.25in} \=
Five -- and two -- colourings of the plane lattice with
four -- rotation axis $C_4$. \\
\>            \>
The first case $n=5$ does not possess
the permutation invariance of five sublattices, \\
\>            \>
which have not had equal fractions in 
the unit cell. In opposite, the second case \\
\>            \>
$n=2$ does possess both these properties.
\end{tabbing}
\begin{tabbing}
\= FIG.2. \hspace{.25in} \=
Two -- and three -- colourings of the plane lattice
with modular sublattices, which\\
\>            \>
preserve the $C_3$ rotation axis. Both lattices
are built out of sublattices which \\
\>            \>
preserved the full permutation
invariance of the colours of equal fractions.
\end{tabbing}

\end{document}